\documentclass[5p]{elsarticle}
\usepackage{amssymb}
\usepackage{graphicx}
\usepackage{lineno,hyperref}
\modulolinenumbers[5]
\def\gr{$\gamma$-ray}

\journal{JAstroparticle Physics}

\begin{document}

\begin{frontmatter}

\title{Evidence  the Galactic contribution to the IceCube astrophysical neutrino flux}
 \author{Andrii Neronov$^{1}$}
 \author{Dmitry Semikoz$^{2}$}
\address{$^{1}$ISDC, Astronomy Department, University of Geneva, Ch.~d'Ecogia 16, Versoix 1290, Switzerland}
\address{$^{2}$AstroParticle and Cosmology (APC), 10 rue Alice Domon et L\'eonie Duquet, F-75205 Paris Cedex 13, France}
\begin{abstract}
We show that the Galactic latitude distribution of IceCube astrophysical neutrino events with energies above 100~TeV is inconsistent with the isotropic model of the astrophysical neutrino flux. Namely, the Galactic latitude distribution of the events shows an excess at low latitudes $|b|<10^\circ$ and a deficit at high Galactic latitude $|b|\gtrsim 50^\circ$. We use Monte-Carlo simulations to show that the inconsistency of the isotropic signal model with the data is at $\gtrsim 3\sigma$ level, after the account of trial factors related to the choice of the low-energy threshold and Galactic latitude binning in our analysis.
\end{abstract}


\end{frontmatter}

\section{Introduction}

The discovery of astrophysical neutrino signal by IceCube experiment has started a new field of neutrino astronomy  \cite{IceCube_PeV,IceCube_1yr,IceCube_3yr,IceCube_2yr,IceCube_combined_2015,IceCube_muonnu_2015}. 
The signal is a relatively soft (i.e. softer than $dN_\nu/dE\propto E^{-2}$) powerlaw. Recent analysis of the data \cite{IceCube_combined_2015} excludes the $1/E^2$ type spectrum at more the 3.8 $\sigma$ level. 
The  signal is consistent with a powerlaw  $dN_\nu/dE=A\left(E/100\mbox{ TeV}\right)^{-\Gamma_{\nu}}$ with normalisation $A=6.7^{+1.1}_{-1.2}\times 10^{-18}$~(GeV cm$^2$ s sr)$^{-1}$ and slope $\Gamma_\nu= 2.50\pm 0.09$ \cite{IceCube_combined_2015}.  

 It is not clear yet what kind of astronomical source(s) are detected by IceCube.   The overall distribution of the three-year event sample is consistent with an isotropic distribution \cite{IceCube_3yr}, while a separate fit of the Northern and Southern hemisphere signals in the four-year signal shows a preference to a harder spectrum in the Northern hemisphere \cite{IceCube_combined_2015} which could potentially be due to the presence of a softer  contribution of the flux from the inner Galaxy  in the Southern hemisphere \cite{Neronov:2014uma,tchernin13a}. Otherwise, the overall approximate isotropy of the signal would point to its extragalactic origin.  
 
The unexpected properties of the observed signal challenge pre-existing theoretical models. Softness of the spectrum disfavours a range of models of extragalactic sources in which the neutrino flux is generated via interactions of high-energy protons with soft photon backgrounds. In such models charged pion production and decay which results in neutrino emission is characterised by a relatively high energy threshold. Neutrino energies in the 10-100~TeV  range are typically  much below the threshold and the neutrino spectrum in this energy range is expected to be much harder than observed \cite{stecker91,mannheim92,neronov02,tchernin13,tkachev}. 

Models which are favoured by the data are those in which the neutrino flux is produced in proton-proton (or, more generally, nuclei-nuclei) interactions. In this case, the energy threhshold for the pion production is as low as $\sim 100$~MeV. The neutrino spectrum above this threshold approximately repeats the parent proton / nuclei spectrum \cite{kelner06}. Measurement of $\Gamma_\nu\simeq 2.5$ just tells that the parent proton  / nuclei spectrum is a powerlaw with the slope $\Gamma_p\simeq \Gamma_\nu\simeq 2.5$ \cite{Neronov:2014uma}. The parent protons / nuclei with such spectrum could be cosmic rays residing in the Milky Way galaxy \cite{Neronov:2014uma,Neronov:2013lza,neronov15a,bykov15,ahlers15,ahlers14} (the Galactic component of neutrino flux). Otherwise, if the average spectrum of Galactic cosmic rays is much softer, the Galactic neutrino flux contribution to the IceCube signal is expected to be negligible \cite{kachelriess14,ahlers14} and neutrinos have to originate from extragalactic sources, like  star forming galaxies \cite{murase1,murase14} or radio galaxies / BL Lacs \cite{tchernin13,Giacinti:2015pya} (the extragalactic component of the flux). 

The extragalactic component of the neutrino flux is rather strongly constrained by the measurements of the isotropic \gr\ background (IGRB) \cite{murase1,Neronov:2014uma}. The IGRB spectrum is a powerlaw with the slope comparable to that of the neutrino spectrum, $\Gamma_\gamma\simeq 2.4$, but with normalisation which is approximately an order of magnitude lower than that of the neutrino spectrum \cite{fermi_IGRB,fermi_IGRB1}. At the same time, the fluxes of \gr s and neutrino prouduced in $pp$ interactions are expected to have approximately equal both normalisations and slopes of the  spectra \cite{kelner06}. The IGRB constraint could be avoided if one assumes that the slope of the neutrino spectrum varies with energy, i.e. if the neutrino spectrum gets harder below $\simeq 1$~TeV.  The broken powerlaw spectrum hardening at low energies is naturally expected in a situation when the low energy protons are trapped in the source (or its host galaxy), while protons with energies higher than a certain threshold $E_\tau$ escape. This  could be the case, e.g. for protons accelerated in the central engines and/or jets of radio loud active galactic nuclei  \cite{Giacinti:2015pya} . The slope of the neutrino spectrum changes around $(0.01..0.1)E_\tau$ \cite{kelner06}. Below this energy it follows original proton spectrum, which has power law index $2.1-2.2$ for usual Fermi acceleration models,  while at higher energies diffusion of protons from the source through the turbulent magnetic field with Kolmogorov turbulence spectrum is expected to soften the spectrum to $2.5$~\cite{Giacinti:2015pya}.   

The possibility of non-negligible Galactic contribution is indicated by the consistency of the all-sky \gr\ and neutrino spectra, which follow the same powerlaw over some five decades in energy (from 10~GeV up to PeV)~\cite{Neronov:2014uma}. The \gr\ all-sky spectrum is dominated by the Galactic contribution, so that it is natural to expect that the Galactic component is also present in the neutrino flux. The analysis of Ref. \cite{IceCube_3yr} has searched for the correlation of the arrival directions of neutrinos with energies above 30~TeV in the three-year data set of IceCube. This analysis has found that the best correlation is at the level of $\simeq 2.5\sigma$ pre-trial in the angle $\pm 7.5^\circ$ around the Galactic Plane and at the $\simeq 2.2\sigma$ level (2.8\% chance coincidence probability) after the trial factor is taken into account.

Below we demonstrate that the neutrino four-year IceCube signal in the energy band  above $100$ TeV \cite{IceCube_3yr,IceCube_ICRC15}, which is free from the residual atmospheric neutrino and muon background \cite{IceCube_3yr}, shows an evidence for the Galactic component.

\section{Anisotropy properties of neutrino signal}

\begin{figure}
\includegraphics[width=\linewidth]{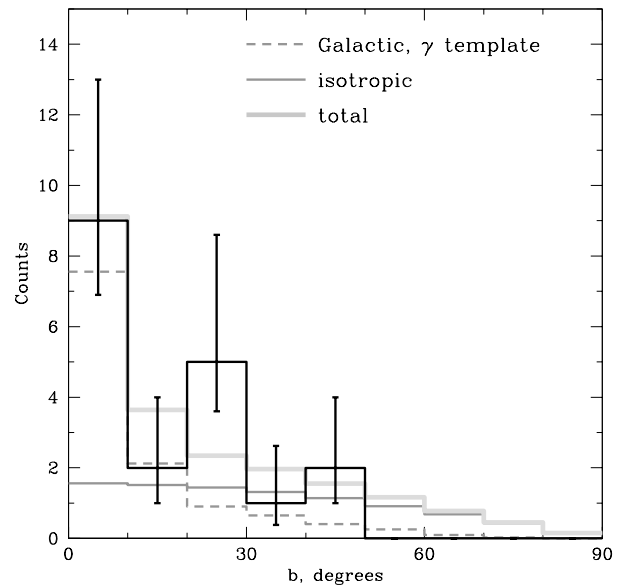}
\caption{Galactic latitude profile of the $E>100$~TeV IceCube neutrino signal. Dark grey solid  histogram shows the expected profile of the isotropic neutrino signal. Dashed dark grey histogram shows the Galactic component profile. Thick light grey solid histogram shows the sum of the Galactic and extragalactic components.    }
\label{fig:profile}
\end{figure}

The Galactic and extragalactic contributions to the neutrino flux could be distinguished based on the difference in the expected distribution of the signal over the sky. The extragalactic flux should be isotropic, while the Galactic flux should show anisotropy toward the Galactic Plane, where most of the target material for the cosmic ray interactions is found. Low statistics of the neutrino signal and uncertainties in the modelling of the Galactic neutrino flux prevented a sensible analysis which would give definitive conclusions on the presence of the Galactic and extragalactic contributions in the first three years of IceCube data \cite{Neronov:2014uma,ahlers15}. The overall distribution of neutrino signal on the sky in the energy band above 30 TeV is consistent with an isotropic distribution \cite{IceCube_3yr,IceCube_combined_2015}, i.e. with the extragalactic signal. However, at 30~TeV the IceCube signal still has a significant contribution from the atmospheric neutrino and muon background which could dilute the weak anisotropy signal. 

A more clean anisotropy analysis could be performed in the energy band above 100~TeV, where the signal is almost backgorund-free. The updated results of 4-year IceCube exposure show 19 events in this energy band with only one background \cite{IceCube_combined_2015,IceCube_ICRC15}.  Fig. \ref{fig:profile} shows the distribution of the detected $E>100$~TeV neutrino events in Galactic latitude. One could notice two features in this distribution. First, the low Galactic latitude bin $|b|<b_{low}=10^\circ$ contains a large number of events (9 out of 19). Next, the bins at high Galactic latitude (above $b_{high}=50^\circ$)  contain no events at all. 

To some extent, the lower number of counts in the bins at high Galactic latitude could be attributed to the smaller solid angle spanned by these bins. To verify if this would provide a satisfactory explanation of the deficit of neutrino counts at high Galactic latitudes, we have performed  Monte-Carlo (MC) simulations of the expected sky distribution of the neutrino signal. The MC simulation takes into account of the declination dependence of the IceCube effective area, derived from the information reported in the Ref. \cite{IceCube_3yr}. This declination dependence leads to a difference in the effective exposure in different Galactic latitude bins. The MC simulation generates the number of events proportional to the exposure in each declination bin.  The events are randomly distributed in the Right Accention. The MC events are then re-mapped in the Galactic coordinates. 

The Galactic latitude distribution of events expected in the isotropic flux model found from MC simulations is shown with the dark grey solid line histogram in Fig. \ref{fig:profile}. If the isotropic flux is normalised on the total number of events, the isotropic model predicts 4.6 events at $|b|>50^\circ$. The probability to find no events in this latitude range is  $p_{|b|>50^\circ}=5\times 10^{-3}$. 

The isotropic model is also inconsistent with the low Galactic latitude data, which shows an excess over the data.  The tension between the model and the high Galactic latitude / low Galactic latitude data could be characterised in a quantitative way using the MC simulations, which show that the probability to find simultaneously the 0 counts at $|b|>50^\circ$ and $\ge 9$ counts in the $|b|<10^\circ$ bin is $10^{-5}$. This corresponds to a $4.4\sigma$ level inconsistency between the model and the data in terms of equivalent Gaussian statistics signal.  

The tension with the data could be readily removed via addition of the Galactic component of the flux. Following Ref. \cite{Neronov:2014uma}, we model this Galactic contribution based on the $E>300$~GeV \gr\ data of Fermi/LAT. The Galactic component, convolved with the IceCube point spread function derived from the distributions of angular uncertainties of detected IceCube events,  is shown by the dashed histogram in Fig. 1. One could see that the Galactic component explains the excess in the first bin and it gives a negligible contribution at high Galactic latitude. The neutrino flux model which contains only Galactic component  is also inconsistent with the data, but at a lower significance level ($\simeq 2\sigma$). A model which contains 50\% contributions from the Galactic and extragalactic components provides a satisfactory fit to the data (Fig. 1). 

\section{Account of the energy and angular binning trail factors}

The above conclusion on the $4.4\sigma$ level  inconsistency of the isotropic model with the data adopts certain choice of the energy range and angular binning for the calculation of probabilities. To have a fair judgement of significance of this inconsistency, one needs to take into account the trial factor related to the choice of the Galactic latitude bins where the inconsistency of the model with the data is largest and to the restriction of the analysis to the events above a particular energy. 

There is no trial factor to be taken into account for the general choice of the pattern "high Galactic Latitude \& low Galactic latitude"  bins.  Indeed, the highest / lowest Galactic latitude bins are those  which contain the lowest / highest  Galactic signal. Any presence of the Galactic signal would reveal itself at the first place in these Galactic latitude bins as a deficit / excess of the data over the model.  

Typical width of the "low Galactic latitude" excess of the Galactic neutrino signal is determined by (a) the angular resolution of the IceCube telescope (which is about $\theta_{PSF}\simeq 10^\circ$ for the "cascade"type events \cite{IceCube_PeV,IceCube_3yr}) and (b) by the angular width of the Galactic Plane itself. 

The width of the Galactic Plane depends on the direction on the sky. This "width" could be defined based on the \gr\ observations of the diffuse Galactic emission reported in the Ref. \cite{fermi_diffuse_2012}. Defining it as the angular distance from the Galactic Plane at which the \gr\ signal drops by a factor of 2, one could find that in almost all directions, the angular span of the Galactic Plane is within $|b|\lesssim 10^\circ$. This could be readily understood using the following qualitative argument.  Most of the target material for cosmic ray interactions in the Galactic Plane is concentrated in a narrow disk of the scale height $H\sim 200$~pc \cite{MW_disk,MW_disk1}. The density of the Disk drops beyond $R\sim 10$~kpc \cite{MW_disk,MW_disk1}, This disk, observed from the position of the Solar system has different angular span in different directions. The estimate of the  angular width of the Disk in the direction of the Outer Galaxy,  $\Theta\lesssim H/(R-R_\odot)\simeq 5^\circ$ is consistent with the observed $\Theta\sim 10^\circ$ extent of the pion decay \gr\ signal in Galactic latitude in the outer Galaxy $90^\circ<l<270^\circ$ observed by Fermi/LAT \cite{fermi_diffuse_2012}. The angular span of the disk is smaller in the inner Galaxy region, shrinking to $H/R_\odot\sim 1.5^\circ$ in the direction of the Galactic Centre \cite{fermi_diffuse_2012}. 

Overall, the angular width of the Galactic component of the neutrino signal is expected to be somewhat larger than the average width of the Galactic Plane, $10^\circ$. This is illustrated in Fig. \ref{fig:profile} which shows the Galactic latitude profile of the Galactic component calculated using the \gr\ template convolved with the IceCube point spread function for the cascade events. 

\begin{figure}
\includegraphics[width=\linewidth]{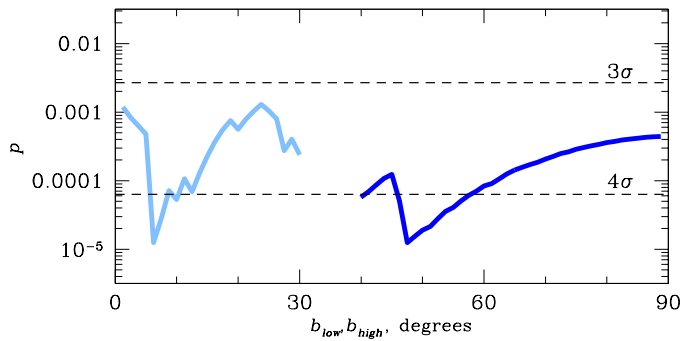}
\caption{Top: dependence of the chance coincidence probability for the isotropic model to describe the $E>100$~TeV data as a function of the cut defining the "low Galactic latitude" region $b_{low}$ (light blue curve on the left) and  as a function of the cut defining the "high Galactic latitude" region $b_{high}$ (dark blue curve on the right).}
\label{fig:probability}
\end{figure}

Light blue color curve on the left of  Fig. \ref{fig:probability} shows the dependence of the chance coincidence probability $p$ on $b_{low}$. One could see that the probability is minimised to $p\simeq 10^{-5}$  at $b_{low}=6.25^\circ$, which is comparable to the expected width of the Galactic component of the neutrino signal calculated based on the \gr\ template.

Dark blue color curve on the right of Fig. \ref{fig:probability} shows the level of inconsistency of the isotropic model with the data as a function of the $b_{high}$ parameter. One could see that the chance coincidence probability reaches a broad minimum in the range $30^\circ<b_{high}<60^\circ$. 

o take into account the trial factor introduced by choosing the values of $b_{low}$, $b_{high}$ at which the inconsistency of the isotropic model with the data is maximal, we allow $b_{low}$ and $b_{high}$ to vary in the limits $0<b_{low}<b<30^\circ$ (as in the Ref. \cite{IceCube_3yr}) and $40^\circ<b_{high}<90^\circ$. We estimate the post-trial probability of the isotropic model to describe the data by counting the number of Monte-Carlo simulated data sets in which the excess of event counts at $|b|<b_{low}$ and deficit of events at $|b|>b_{high}$ leads to the chance coincidence probability $\le 10^{-5}$ . This results in the post-trial probability $p=2\times 10^{-4}$, or a $3.7\sigma$ level inconsistency of the model with the data. 

The inconsistency of the spatial distribution of the neutrino signal with the isotropic model is expected to decrease with the decrease of the energy threshold of the analysis. Indeed, at lower energies, the astrophysical signal gets diluted by the increasingly important atmospheric muon and neutrino backgrounds. The background components obviously do not correlate with the Galactic Plane and their Galactic latitude distribution is similar to that of the isotropic astrophysical flux component. Thus, the deviations from the isotropic distribution are more difficult to detect at low energies.

\begin{figure}
\includegraphics[width=\linewidth]{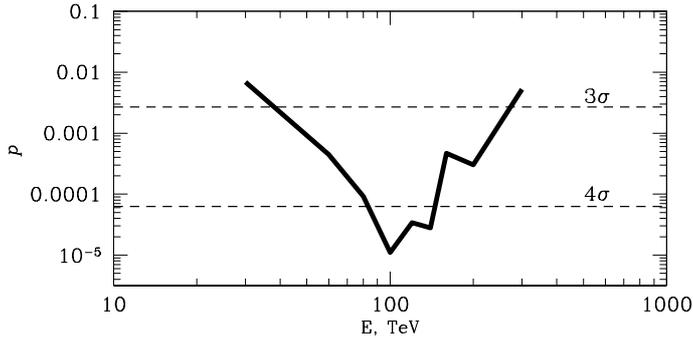}
\caption{Dependence of the chance coincidence probability for the isotropic model to describe the data as a function of the low energy threshold.}
\label{fig:probability_energy}
\end{figure}

Fig. \ref{fig:probability_energy} shows the dependence of the level of inconsistency of the isotropic flux model with the data as a function of the low energy threshold. One could see that the chance coincidence probability reaches a minimum at the threshold $E=100$~TeV. At lower energies the probability rapidly increases because of the increase of the background. At higher energies the probability increases because of the decrease of the signal statistics.

We estimate the trial factor due to the particular choice of the energy threshold by extending the analysis down to 30~TeV (the low energy threshold of the reported IceCube data \cite{IceCube_3yr,IceCube_ICRC15}). The number of generated Monte-Carlo events of different energies follows the energy distribution of the really observed events. The post-trial probability for the isotropic model to describe the data is calculated by counting the number of occurrences of the low Galactic latitude excesses and high Galactic latitude deficits leading to the inconsistency level $p\le 10^{-5}$ in the simulated data sets after the adjustment of the energy threshold to the value which minimises the probability for the isotropic model to describe the data.  This results in the post-trial probability $p=1.7\times 10^{-3}$, i.e. to the inconsistency of the isotropic model with the data at $3.1\sigma$ level. 

\section{Comparison with the previous search of the Galactic component}

Search for the excess of neutrino signal in the direction of the Galactic Plane was previously reported in Ref. \cite{IceCube_3yr} based on the analysis of the three-year dataset of IceCube. The difference between our approach and that of the Ref. \cite{IceCube_3yr} is 
\begin{itemize}
\item the account of not only the low Galactic latitude excess, but also of the high Galactic latitude deficit, characteristic for the Galactic flux component and
\item dedicated analysis of the high-energy part of the  neutrino spectrum.
\end{itemize}
As it is explained above, both modifications potentially lead to an increase of the sensitivity of the search via more precise account of the signal morphology and via suppression of the unrelated background of atmospheric muons and neutrinos. 

To assess the improvement of sensitivity due to these modifications, we have repeated the analysis restricting the data set to the three-year IceCube data of Ref. \cite{IceCube_3yr}. We find that the  chance coincidence probability for the isotropic model to describe the three-year data above 100~TeV energy threshold is $p=1.3\times 10^{-3}$ pre-trial and $p=7.8\times 10^{-2}$ post-trial. This should be compared to the $p\simeq 1$\% / $p=2.8$\% pre-trial / post-trial probability found in Ref. \cite{IceCube_3yr} without an additional verification of the high Galactic latitude deficit, but using the likelihood analysis technique. The optimal width of the Galactic Plane is found to be $b_{low}=6.25^\circ$ (compare to the $7.5^\circ$ found in the Ref. \cite{IceCube_3yr}) and the optmal energy cut is at $E=200$~TeV.

\section{Conclusions}

To summarise,  we have shown that the Galactic latitude distribution of the neutrino events in the 4-year IceCube data set at highest energies $E>100$ TeV is inconsistent with the hypothesis of the isotropic neutrino flux at about $3\sigma$ level. This inconsistency suggests the presence of a significant 
 Galactic contribution to the signal. The statistics of the signal is still to low for the measurement of the relative importance of the Galactic and extragalactic flux components. A broad range of models with sizeable (the same order of magnitude) contributions of the two components are consistent with the present data.  Taking the 3$\sigma$ evidence level found in the 4-year data sample as an indication, one could estimate that the final verification of the presence of the Galactic component of the neutrino flux at $E>100$~TeV would require a decade-long IceCube exposure. A much faster verification of the Galactic signal will be possible with the IceCube Generation 2 detector \cite{IceCube_gen2}. Northern hemisphere neutrino telescopes, like km3net, should be able to detect the Galactic component of the signal at much lower energies due to the lower atmospheric background in the up-going muon neutrino channel. The results of analysis of the data of ANTARES  neutrino telescope show that already the sensitivity of this telescope is marginally reaching the expected flux level of the Galactic component (provided that it constitutes 100\% of the IceCube signal) \cite{Antares_ICRC15}. This shows that a detailed study of the spectral and morphological properties of the Galactic signal will be possible with the km$^{3}$ scale detector, km3net \cite{km3net}. 

\section*{Acknowledgement}

We would like to thank F.Halzen for the discussion of the results. We are also grateful to the anonymous referee for useful comments and suggestions which have significantly improved the quality of the reported analysis. 

\section*{References}

\bibliography{IceCube_4yr_AstropPhys}

\end{document}